\documentclass{article}

\usepackage{varioref}
\usepackage{amsmath,amsfonts,amsthm,mathrsfs}
\usepackage[normalem]{ulem}
\usepackage{multirow,color,graphics}
\usepackage{amsmath}
\usepackage{amsfonts}
\usepackage{amssymb}
\usepackage{jheppub}
\usepackage{enumerate}
\usepackage{fancyvrb}
\usepackage{verbatim}
\usepackage{wrapfig}
\usepackage{appendix}
\usepackage{amstext}
\usepackage{amssymb}
\usepackage{graphicx}
\usepackage{color}
\usepackage{varioref}
\usepackage{multirow,graphics}
\usepackage{epstopdf}
\newcommand{\nn}{\nonumber}

\numberwithin{equation}{section}

    \newcommand{\beq}{\begin{equation}}
    \newcommand{\eeq}{\end{equation}}
    \newcommand\beqa{\begin{eqnarray}}
    \newcommand\eeqa{\end{eqnarray}}
\newcommand\bea{\begin{array}}
\newcommand\eea{\end{array}}

\newcommand{\bQ}{{\bf Q}}
\newcommand{\bP}{{\bf P}}

\newcommand{\cQ}{{\cal Q}}

\newcommand{\la}[1]{\label{#1}}
\newcommand{\eq}[1]{(\ref{#1})}

    \def\CP{{\cal P}}
    
    \def\CR{{\cal R}}

    \def\bQ{{\bf Q}}
    \def\bP{{\bf P}}
    
    \def\bQt{\tilde{\bf Q}}

\makeatletter
     \@ifundefined{usebibtex}{} {}
\makeatother

\preprint{~LPTENS-15/02 \\ \rightline{~IPhT-t14/235}}

    \def\bQ{{\bf Q}}

        \def\bP{{\bf P}}

     \def\Tr{\text{Tr}}
     \def\l{\lambda}

\newcommand{\gQ}{{\cal Q}}

\title{\Large QCD Pomeron from AdS/CFT Quantum Spectral Curve }
\author[a,b]{~~Mikhail Alfimov}
\author[c,d]{~~Nikolay Gromov}
\author[a,e]{~~~Vladimir Kazakov}

\affiliation[a]{LPT, \'Ecole Normale Superieure, 24, rue Lhomond 75005
  Paris, France}
\affiliation[b]{ Institut de Physique Theorique, Orme des Merisiers, CEA Saclay 91191 Gif-sur-Yvette Cedex, France \& P.N. Lebedev Physical Institute, Leninskiy pr. 53, 119991 Moscow, Russia \& Moscow Institute of Physics and Technology, Institutskiy per. 9, 141700 Dolgoprudny, Russia }
\affiliation[c]{Mathematics Department, King's College London,
The Strand, London WC2R 2LS, UK.}
\affiliation[d]{St.Petersburg INP, Gatchina, 188 300, St.Petersburg,
  Russia}
\affiliation[e]{Universit\'e Paris-VI, Place Jussieu, 75005 Paris,
  France}

\abstract{
Using the methods of the recently proposed Quantum Spectral Curve (QSC)
originating from integrability of ${\cal N}=4$ Super--Yang-Mills theory we analytically continue the
scaling dimensions of  twist-2 operators
and reproduce
the so-called pomeron eigenvalue of the Balitsky-Fadin-Kuraev-Lipatov (BFKL) equation.
Furthermore, we recovered the Faddeev-Korchemsky Baxter equation for Lipatov's spin chain
and also found its generalization for the next-to-leading order in the BFKL scaling.
Our results provide a non-trivial test of QSC describing the exact spectrum in planar ${\cal N}=4$ SYM
at
 infinitely many loops for a highly nontrivial non-BPS quantity
and also opens a way for a systematic expansion in the BFKL regime.
}

\begin{document}



\maketitle
{\addtocounter{page}1}


\section{Introduction}

The exact spectrum of anomalous dimensions in the planar
\({\cal N}=4\) SYM theory is described by the recently proposed in \cite{Gromov:2013pga,Gromov:2014caa} Quantum
Spectral Curve (QSC) equations, following a long and successful study of this problem
during the last decade \cite{Beisert:2010jr}.  The paper \cite{Gromov:2014caa}
generalizes  \cite{Gromov:2013pga} to an arbitrary operator/state of the
theory and reveals its general mathematical structure in terms of the analytic Q-system.
The QSC approach has already a history of a few non-trivial tests  and  applications.   In the  weak coupling limit,  the one-loop dimensions for twist-\(L\) operators of the type \(\Tr (Z\nabla^SZ^{L-1})\) of  \(\mathfrak{sl}_2\)  sector were reproduced    \cite{Gromov:2013pga}  and then the method was applied for calculating the dimension of Konishi operator at 10 loops \cite{Marboe:2014gma}. For the small \(S\) expansion of anomalous dimension of twist-2 operator  \(\gamma=f_1(\lambda)S+f_2(\lambda)S^2+{\cal O}(S^3)\), the  slope function \(f_1\) \cite{Basso:2011rs}, exact at any 't Hooft coupling $\lambda$, was reproduced from QSC and the ``curvature" function \(f_2\) was then found in \cite{Gromov:2014bva}.  The  results for the cusp anomalous dimension at small angle of the cusp, known from localization \cite{Correa:2012at} and TBA \cite{Correa:2012hh,Gromov:2012eu,Gromov:2014bva}, we reproduced in an elementary way from QSC in \cite{Gromov:2013pga}. The QSC method was recently generalized to the case of ABJM theory \cite{Cavaglia:2014exa} which allowed the efficient calculation of the ABJM slope function and helped to identify the mysterious interpolating function fixing the dependence  of dispersion relation on the `t Hooft coupling \(\lambda\)  \cite{Gromov:2014eha}\footnote{recently tested by a heroic strong coupling two loop calculation in \cite{Bianchi:2014ada}}
and given the last missing element in the solution of the spectral problem for this model.

Here we demonstrate another application of the QSC to an important problem -- the calculation of conformal dimensions \(\Delta\) of  the twist-2 operators of a type \(\Tr (Z\nabla_+^SZ)\) belonging to the \(\mathfrak{sl}_2\) sector in the BFKL limit, corresponding to a double scaling regime of small 't~Hooft constant \(g\equiv \sqrt \l /(4\pi)\to0\) and the Lorentz spin \(S\) approaching to \(-1\), whereas   the ratio \(\Lambda\equiv \frac{g^2}{S+1}\) is kept fixed. We  will  reproduce the
famous formula for this dimension,
obtained in \cite{Jaroszewicz:1982gr,Lipatov:1985uk,Kotikov:2002ab} from the direct re-summation of Feynman graphs:
\begin{eqnarray}\label{KL}
\frac{1}{4\Lambda}=-\psi\left(\frac{1}{2}-\frac{\Delta}{2}\right)-\psi\left(\frac{1}{2}+\frac{\Delta}{2}\right)+2\psi(1)+{\cal O}(g^2)
\label{BFKLdimension}
\end{eqnarray}
where
\(\psi(x)=\frac{\Gamma'(x)}{\Gamma(x)}\). Remarkably, this  result is also known to be valid for the  pure Yang-Mills theory in the planar limit since only the gluons appear inside the Feynman diagrams of \({\cal N}=4\) SYM at LO! In \cite{Fadin:2011jg} the conformal invariance of the BFKL kernel with the characteristic function \eqref{BFKLdimension} was shown. This formula was a result of a long and remarkable history of applications of the BFKL method to the study of Regge limit of high energy scattering amplitudes and correlators in QCD \cite{Kuraev1976c,Kuraev1977,Balitsky1978,Balitsky1996,Faddeev:1994zg} and in the \({\cal N}=4\) SYM theory \cite{Kotikov:2002ab,Kotikov:2007cy,Balitsky:2009yp,Balitsky2013a,Balitsky2009,Cornalba:2008qf,Cornalba:2009ax,Costa:2013zra,Costa:2012cb}. The effective action for the high-energy processes in nonabelian gauge theories was derived in \cite{Lipatov:1995pn}. 
Recently, certain scattering amplitudes describing the adjoint sector (single reggeized gluon) were computed by means of all loop integrability in the BFKL limit \cite{Basso2014g} in the
integrable polygonal Wilson loop formalism \cite{Basso:2013vsa}.

To recover the formula \eqref{BFKLdimension} from our QSC approach, we will have to compute certain quantities not only in the LO, but also in the NLO. In particular, we extract from the analytic Q-system describing QSC
the Baxter-Faddeev-Korchemsky equation for the pomeron wave function  \cite{Lipatov:1993yb,Faddeev:1994zg,DeVega:2001pu,Derkachov:2002pb,Derkachov:2002wz,Derkachov:2001yn} in the LO and generalize it to the NLO.
Some other ingredients of the QSC, entering the underlying so called \(\bP\mu\) and \(\bQ\omega\) equations,   will be determined in the LO or even up to NLO.
 These calculations
lay out a  good basis for the construction of a systematic BFKL expansion of this anomalous dimension in planar \({\cal N}=4\) SYM,  now known only up to  NLO correction to \eqref{KL} from the direct computation of  \cite{Fadin:1998py,Kotikov:2000pm,Kotikov:2002ab}.

Our method, designed here for the case of pomeron singularity (a bound state of two reggeized gluons)  should be applicable to the study of a bound state of \(L\) reggeized gluons as well.

Let us stress that the main result of this paper -- the correct reproduction of this formula from the QSC --  is a very non-trivial test of the QSC
 as well as of the whole integrability approach to planar AdS/CFT spectrum.  It sums up an infinite
 number of the so-called wrapping corrections \cite{Bajnok:2008qj,Lukowski:2009ce} providing a test for
 infinitely many loops for a highly nontrivial non-BPS quantity.

\section{Motivation from Weak Coupling}\label{par_weak_coupling}
To get an idea of the QSC it is instructive to start from the example of a non-compact, $sl_2$ Heisenberg spin chain with negative values of spins $-s$ (usually denoted as XXX${}_{-s}$), where \(s\) is not necessarily integer or half-integer.
In particular the Heisenberg spin chain with $s=1/2$ describes the twist-2 operators at weak coupling at one loop. Furthermore, we will see that it is also responsible for the BFKL regime of these operators.
Let us consider the case with two particles (two nodes of the spin chain) for simplicity.
For this integrable model
the problem of finding its spectrum reduces to the Baxter equation
\beq\la{Baxter}
T(u)Q(u)+(u-is)^2Q(u-i)+(u+is)^2Q(u+i)=0\;,
\eeq
where $T(u)$ is some $2^{\rm nd}$ order polynomial
which encodes the total spin $S$ of the
state via $T(u)=-2u^2+\left[S^2-S+4sS+2s^2\right]$ for zero momentum states.
For the case of the one-loop spectrum of the twist-2 operators $S$ corresponds to the operator with $S$ covariant derivatives.

This equation is in many respects similar to the usual Schr\"odinger equation, where $Q(u)$
plays the role of a wave function and $T(u)$ is an external potential.
When $S$ is integer one can find a polynomial solution of \eq{Baxter} of degree $S$, which we denote  as $Q_1(u)$.
The energy of the state is then given  by
\beq\la{energy}
\Delta=2+S+2ig^2\left. \partial_u\log\frac{Q_1(u+i s)}{Q_1(u-i s)}\right|_{u=0}\;.
\eeq
For $s=1/2$ this polynomial can be found explicitly  (see, for example, \cite{Eden:2006rx}) and it gives for the energy
\beq\la{oneloopres}
\Delta(S)=2+S-8g^2\sum_{n=1}^S \frac{1}{n}\;.
\eeq
An important point is that since \eq{Baxter} is a second degree finite difference equation there must be another solution to it. It is easy to show, by plugging \(u^\alpha\) into the equation and taking \(u\to\infty\), that
the second linearly independent solution, denoted as \(Q_2(u)\) has the asymptotics \(u^{1-4s-S}\). It is clear from here that for \(s>0\) both solutions cannot be polynomial. Unavoidably, \(Q_2(u)\) should have infinitely many zeros and poles.
To see the positions of these poles we build the Wronskian $Q_{12}(u)$ out of these two solutions
\beq\la{QQtoy}
Q_{12}=Q_1^+Q_2^--Q_1^-Q_2^+.
\eeq
 As a consequence of \eq{Baxter} it satisfies $
\frac{Q_{12}^+}{Q_{12}^-}=\frac{(u- is)^2}{(u+ is)^2}
$
which can be solved to give
\beq\la{Q12}
Q_{12}=\left(\frac{\Gamma(-iu+1/2-s)}{\Gamma(-iu+1/2+s)}\right)^2\;
\eeq
from where we see explicitly that $Q_{12}$ has second order poles at $u=i(n-s+\tfrac 12),\;n\in{\mathbb Z_{\geq 0}}$ and is analytic in the upper half plain.
We also see that $Q_2$ should have double poles at $u=-i(n-s+1)$.

The normalization in \eq{energy}
is chosen so that for $s=1/2$ it gives precisely one-loop dimensions of twist
two operators with $S$ covariant derivatives and two scalars. To pass to the BFKL regime
and take the limit $S\to -1$ we have to analytically continue away from even integer $S$.
The analytic continuation of the energy itself $\Delta(S)$ is naturally given by the following rewriting of \eq{energy}
\beq\la{EnergySS}
\Delta(S)=2+S-8g^2 \sum_{n=1}^\infty \left(\frac{1}{n+S}-\frac{1}{n}\right)\;.
\eeq
In this form the sum is meaningful for non-integer $S$ and we also  clearly see a pole in this energy at $w=S+1\to 0$:
\beq
\Delta(S)\simeq -\frac{8g^2}{w}\;
\eeq
which reproduces the BFKL prediction \eq{KL} at one loop (which can be found from \eq{KL} by inverting the series for the expansion $\Delta\to 1$).
However, at the level of $Q$-functions it is not immediately clear how to make the analytic continuation.
Indeed, $Q_1$ as  a solution with $u^S$ asymptotics could no longer be polynomial and must also have poles.
Requiring the power-like asymptotics the best it is possible to achieve is to cancel the second order pole and build a unique, up to a constant multiplier,
$Q_1$ so that it has only simple poles.
The singularities of the both ``big" solution $Q_1$ and ``small" solution $Q_2$ are located at $u=-i n-i/2$ for all positive integers $n$  as we discussed before.
These poles result in infinities in the expression for the energy \eq{energy}.
The way to avoid such infinities is to form a regular combination, still solving the Baxter equation,
\beq\label{Qpresc}
Q_1(u)+\cosh(2\pi u) Q_2(u)
\eeq
where all poles are canceled, having however an exponential asymptotics.
One can show that there is a unique, up to an overall normalization,
combination of this form which is regular everywhere in the whole complex plain \cite{Janik:2013nqa,GKunpublished}.
It automatically gives the correct analytic continuation for the dimension \eq{EnergySS}.
In other words, one should find a regular solution of \eq{Baxter}
with the  large positive \(u\) asymptotics $u^S+{\rm const} \; e^{2\pi u} u^{-1-S}$ and plug it into \eq{energy}
to get the correct analytic continuation to non-integer $S$.
We will see how a similar prescription allows to define the QSC for non-physical operators for any $S$.

\section{Quantum Spectral Curve -- Generalities}
The QSC gives a generalization of the above construction to all loops.
When we go away from weak coupling regime
we start exploring all other degrees of freedom of the dual super-string in 10D. Thus the full symmetry group
 ${\rm PSU}(2,2|4)$ emerges which simply means that we should consider generalized Baxter equations with $2+2+4=8$ Q-functions with one index, which we denote as ${\bf P}_a,\;a=1,\dots,4$ ($S^5$ part) and
 ${\bf Q}_i,\;i=1,\dots,4$ ($AdS_5$ part).
 Out of them, we can form Wronskians like in \eq{QQtoy} - which give another $8*7/2=28$ Q-functions with two indices, then we can iterate the process several times. In total we  get $2^8$ various Q-functions.

Another effect which happens at finite coupling is that the poles of Q-functions in the lower-half plane,  described above,
resolve into cuts $[-2g,2g]$ (where $g=\sqrt\lambda/4\pi)$.

Finally, we have to  introduce   new objects -- the monodromies \(\mu_{ab}\) and \(\omega_{ij}\) corresponding to the analytic continuation of the functions \(\bP_a\) and \(\bQ_j\)  under these cuts.
They will be given  by equations \eqref{Pt} and \eqref{omegaQ}.

Below we describe in more details this construction
following \cite{Gromov:2014caa}. We also derive some new relation important for the BFKL applications.

\paragraph{Algebraic properties:} The AdS/CFT Q-system is formed by \(2^8\) Q-functions which we denote as  \(\cQ_{A|J}(u)\) where \(A,J\subset{\{1,2,3,4\}}\) are two ordered subsets of indices.
They satisfy the  QQ-relations\footnote{As usually, we will  denote the shifts w.r.t. the  spectral parameter as \(f^{\pm}\equiv f(u\pm i/2)\) and \(f^{[n]}\equiv f(u+in/2)\) }, generalizing \eq{QQtoy}
\begin{subequations}\label{definingQQ}
    \begin{align}
       \label{QQbb}
       \gQ_{A|I}\gQ_{A ab|I} &=\gQ_{A a|I}^{+} \gQ_{A b|I}^{-}-
       \gQ_{A a|I}^{-}
       \gQ_{A b|I}^{+}\,,\\
       \label{QQff}
       \gQ_{A|I}\gQ_{A|I ij} &=\gQ_{A|I i}^{+} \gQ_{A|I j}^{-}-
       \gQ_{A|I i}^{-} \gQ_{A|I j}^{+}\,,\\
       \label{QQbf}
       \gQ_{A a|I}\gQ_{A|I i} &= \gQ_{A a|I i}^{+}\gQ_{A|I}^{-}-
       \gQ_{A|I}^{+} \gQ_{A a|I i}^{-} \,
     \end{align}
\end{subequations}  and reshuffling a pair of individual indices   (small letters \(a,b,i,j\)) we can express all Q-functions through 8 basic ones.
In addition we also impose the constraints\footnote{By \(\emptyset\) we denote the  empty set}
\(\gQ_{\emptyset |\emptyset}=\gQ_{1234|1234}=1\),   the first one being a normalization and the second
can be interpreted as a consequence of unimodularity of the symmetry group \cite{Gromov:2010km}. The Hodge dual
of this Q-system, built out of the  Q-functions defined through the old ones as
\(\gQ^{A|J}\equiv (-1)^{|A|\;|J|}\gQ_{\bar A|\bar J}\)  satisfies the same QQ-relations.
  Here  the bar over a subset means the subset complementary w.r.t. the full~set \(\{1,2,3,4\)\} and  \(|X|\) denotes the number of indexes in $X$.   We use special notations for 16 most important Q-functions mentioned before:        \(\bP_a\equiv Q_{a|\emptyset}\),  \(\bP^a\equiv Q^{a|\emptyset}\),    \(\bQ_j\equiv Q_{\emptyset|j}\) and \(\bQ^j\equiv Q^{\emptyset|j}\), where \(a,j=1,2,3,4\).

One can think of $\bP_a$ (and $\bP^a$) as of  quantum conterparts of the classical quasimomenta describing the $S^5$ part of the string motion,
whereas $\bQ_i$ (and $\bQ^i$) correspond to the $AdS_5$ part.
A nice feature of the Q-system is that any Q-function can be expressed in terms of
$\bP_a$ and $\bP^a$ or, alternatively, in terms of $\bQ_i$ and $\bQ^i$.
Furthermore, the discontinuity relations for $\bP$'s decouple from the rest of the system
and form a closed system of equations, called $\bP\mu$-system, which carries complete information
about the spectrum of the whole $AdS_5\times S^5$ worldsheet sigma model.
Alternatively, one can decouple $\bQ_i$ and $\bQ^i$ from the rest of the system
getting a description more natural for the $AdS$ type of excitations, called $\bQ\omega$-system.
In different situations one or another description could be more convenient, or even a completely new set of basic Q-functions could be chosen to form a closed set of equations.
At the same time one can always pass from one description to another.

\paragraph{From $\bP_a$ to $\bQ_i$.}
Here we present our new result which allows for the direct transition between these two equivalent systems.
We show in the appendix \ref{Der4orderEq} that, as a consequence of the QQ-relations, $\bP$'s and $\bQ$'s are related through the following $4$th order finite-difference equation
\begin{eqnarray}\label{QPeq}
0=\bQ^{[+4]}D_0-\bQ^{[+2]}
\left[D_1-\bP_a^{[+2]}\bP^{a[+4]}D_0\right]+
\frac{1}{2}\bQ
\left[D_2-\bP_a\bP^{a[+2]}D_1+\bP_a\bP^{a[+4]}D_0 \right]+\text{c.c.}
\end{eqnarray}
where
\begin{eqnarray}{}
& D_0=\det\left(\begin{array}{llll}
\bP^{1[+2]} & \bP^{2[+2]} & \bP^{3[+2]} & \bP^{4[+2]} \\
\bP^{1} & \bP^{2} & \bP^{3} & \bP^{4} \\
\bP^{1[-2]} & \bP^{2[-2]} & \bP^{3[-2]} & \bP^{4[-2]} \\
\bP^{1[-4]} & \bP^{2[-4]} & \bP^{3[-4]} & \bP^{4[-4]}
\end{array}\right), \quad
D_1=\det\left(\begin{array}{llll}
\bP^{1[+4]} & \bP^{2[+4]} & \bP^{3[+4]} & \bP^{4[+4]} \\
\bP^{1} & \bP^{2} & \bP^{3} & \bP^{4} \\
\bP^{1[-2]} & \bP^{2[-2]} & \bP^{3[-2]} & \bP^{4[-2]} \\
\bP^{1[-4]} & \bP^{2[-4]} & \bP^{3[-4]} & \bP^{4[-4]}
\end{array}\right), \\
& D_2=\det\left(\begin{array}{llll}
\bP^{1[+4]} & \bP^{2[+4]} & \bP^{3[+4]} & \bP^{4[+4]} \\
\bP^{1[+2]} & \bP^{2[+2]} & \bP^{3[+2]} & \bP^{4[+2]} \\
\bP^{1[-2]} & \bP^{2[-2]} & \bP^{3[-2]} & \bP^{4[-2]} \\
\bP^{1[-4]} & \bP^{2[-4]} & \bP^{3[-4]} & \bP^{4[-4]}
\end{array}\right). \notag
\end{eqnarray}
The four solutions of this equation give  four  functions \(\bQ_j\).
This relation will be useful for us since, whether  as $\bP\mu$-system is simpler at weak coupling,  $\bQ\omega$-system a priori is more suitable
for the $sl_2$ sector to which the twist-2 operators belong.\footnote{
We also note that in a similar way one can derive similar relation \eq{QPeq} with $\bP$ and $\bQ$ exchanged.}

\paragraph{Analytic properties:}
The Q-system is a generic grassmanian algebraic construction, based entirely on the symmetry group.
To  apply it to our particular model we have to complete it by  analyticity properties.
An important analytic feature of    the AdS/CFT Q-functions is that they are multi-valued functions of \(u\),
with  infinitely many Riemann sheets connected by cuts, parallel to \(\mathbb{R}\),  with  fixed quadratic branch-points at \(u\in\pm2 g+i\mathbb{Z}\) or \(u\in\pm
 2 g+i\mathbb{(Z} +\frac{1}{2})\). According to the arguments of \cite{Gromov:2014caa}
 there are no other singularities on the whole Riemann surface of any Q-function.
 The basic $16$ Q-functions $\bQ$ and $\bP$ have particularly nice properties:
$\bP_a$ and $\bP^a$ have only one ``short" cut \(u\in[-2 g,2g]\) on their main, defining sheet of its Riemann surface,
whether $\bQ_j$ and $\bQ^j$ have only one ``long" cut \(u\in(-\infty,-2g]\cup[2g,\infty)\)  on their main sheet. The rest of the Q-functions can be
expressed in terms of
either $8$ $\bP$'s or $8$ $\bQ$'s using QQ-relations.
Depending on this choice we have two equivalent systems of equations described below.
\begin{figure}
\centering{\includegraphics{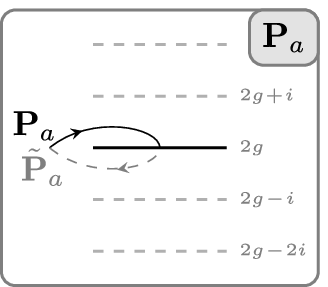}~\includegraphics{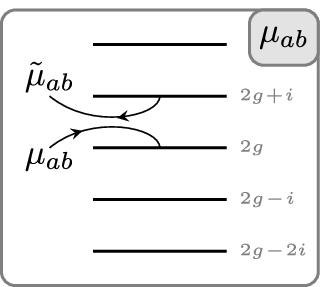}~\includegraphics{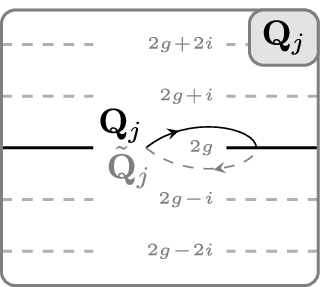}~\includegraphics{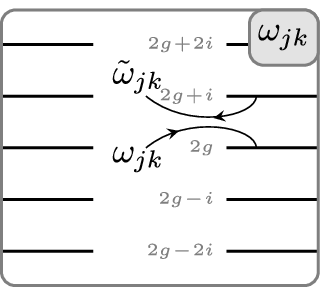}}
 \caption{Cut structure of \(\bP\) and \(\mu\), \(\bQ\) and \(\omega\) and their analytic continuations \(\tilde\bP\) and \(\tilde\mu\), \(\tilde{\bQ}\) and \(\tilde{\omega}\) \cite{Gromov:2013pga,Gromov:2014caa}}
\label{fig:Cuts}
\end{figure}

\paragraph{${\bf P}\mu$-system.}
As we explained above, we can focus on a much smaller closed subsystem constituted of $8$ functions \(\bP_a\) and \(\bP^a\), having only one short cut on the real axis on their defining sheet. To close the system we have to describe their analytic continuation under this cut, to the next sheet, as shown in Fig.\ref{fig:Cuts}. Denoting this continuation by $\tilde \bP$ we simply have \cite{Gromov:2014caa}
\begin{equation}\label{Pt}
\tilde \bP_a=\mu_{ab}(u)\bP^b\;\;,\;\;\tilde \bP^a=\mu^{ab}(u)\bP_b
\end{equation}
where $\mu_{ab}(u)$ is an antisymmetric matrix with unit Pfaffian,
having infinitely many branch points at \(u\in\pm2 g+i\mathbb{Z}\) and \(\mu^{ab}=-\frac{1}{2}\epsilon^{abcd}\mu_{cd}\) is its inverse.
To distinguish between short cut/long cut version of the same function
we add hat/check over the symbol. Then for $\check \mu_{ab}$ we have
the \(i\)-periodicity condition\beq\la{muper}
\check\mu_{ab}(u+i)=\check\mu_{ab}(u)\,.
\eeq
 This means that all cuts are exact copies of each other with the distance \(i\) between them.
The analytic continuation under these cuts is again very simple and is given by \cite{Gromov:2014caa}
\begin{equation}\label{muPPt-eq}
\tilde\mu_{ab}-\mu_{ab}=\bP_a\tilde \bP_b-\bP_b\tilde \bP_a\;.
\end{equation}
Note that if we decide to consider \(\hat\mu_{ab}\) instead of the  periodic \(\check\mu_{ab}\), we can combine the two above  relations into a linear finite difference equation for  \(\hat\mu_{ab}\)
\beq\label{fthorder}
\hat\mu_{ab}^{++}=
\hat\mu_{ab}+\bP_a\hat\mu_{bc}\bP^c-\bP_b\hat\mu_{ac}\bP^c\;.
\eeq
To see this we can take $u$ to be slightly below the real axis,
then ${{\check\mu}}_{ab}(u+i)=\tilde{{\hat\mu}}_{ab}(u)$, as shown in Fig.\ref{fig:Cuts}.

Finally $\bP$'s  satisfy the orthogonality relations  $\bP_a\bP^a=0$ and at large $u$ they should behave as
\beq\la{PAM}
\left(
\bea{c}
\bP_1\\
\bP_2\\
\bP_3\\
\bP_4
\eea
\right)\simeq
\left(
\bea{l}
A_1\; u^\frac{-J_1-J_2+J_3-2}{2}\\
A_2\; u^\frac{-J_1+J_2-J_3}{2}\\
A_3\; u^\frac{+J_1-J_2-J_3-2}{2}\\
A_4\; u^\frac{+J_1+J_2+J_3}{2}
\eea
\right)\;\;,\;\;
\left(
\bea{c}
\bP^1\\
\bP^2\\
\bP^3\\
\bP^4
\eea
\right)\simeq
\left(
\bea{l}
A^1\; u^\frac{+J_1+J_2-J_3}{2}\\
A^2\; u^\frac{+J_1-J_2+J_3-2}{2}\\
A^3\; u^\frac{-J_1+J_2+J_3}{2}\\
A^4\; u^\frac{-J_1-J_2-J_3-2}{2}
\eea
\right)\;.
\eeq
We note, that the coefficients $A_a$ and $A^a$ could be also determined solely in terms of the global symmetry Cartan charges \((S_1,S_2,\Delta|J_1,J_2,J_3)\) of the state,
including the energy \(\Delta\). We will briefly discuss this  below.

\paragraph{$\bQ\omega$-system.}
It may seem that the description in terms of $\bP\mu$-system breaks the symmetry between $AdS_5$ and $S^5$ parts of the string background.
It is possible however to pass to an alternative, equivalent description where the roles of these parts
are interchanged. We will see that we  also have to interchange short and long cuts. To construct this alternative system we can use \eqref{QPeq}
which, for a given $\bP_a$, gives us $4$ linear independent solutions $\bQ_i$ (similarly we construct $\bQ^i$).
Knowing $\bP_a$ and $\bQ_i$ we construct $\cQ_{a|i}$ using \eqref{QPQ}
which allows us to define $\omega_{ij}$
\begin{align}
\label{muomega}
\omega_{ij}=
\cQ_{a|i}^- \cQ_{b|j}^-\mu^{ab} \,.
\qquad
\end{align}
One can show that  $\bQ_a$ defined in this way will have one long cut.
Also $\hat \omega_{ij}$, with short cuts, happens to be periodic $\hat\omega_{ij}^+=\hat\omega_{ij}^-$, similarly to  its counterpart with long cuts  $\check \mu_{ab}$!
Finally, their discontinuities are given by
\begin{equation}\label{tQ=omegaQ}
\tilde\omega_{ij}-\omega_{ij}=\bQ_i\bQt_j-\bQ_j\bQt_i
\end{equation}
\begin{equation}
\label{omegaQ}
\tilde \bQ_i=\omega_{ij} \bQ^j\;.
\end{equation}
Similarly to \eq{PAM}, we have the large \(u\)  asymptotics
\beq\la{QAM}
\left(
\bea{c}
\bQ_1\\
\bQ_2\\
\bQ_3\\
\bQ_4
\eea
\right)\simeq
\left(
\bea{l}
B_1\; u^\frac{+\Delta-S_1-S_2}{2}\\
B_2\; u^\frac{+\Delta+S_1+S_2-2}{2}\\
B_3\; u^\frac{-\Delta-S_1+S_2}{2}\\
B_4\; u^\frac{-\Delta+S_1-S_2-2}{2}
\eea
\right)\;\;,\;\;
\left(
\bea{c}
\bQ^1\\
\bQ^2\\
\bQ^3\\
\bQ^4
\eea
\right)\simeq
\left(
\bea{l}
B^1\; u^\frac{-\Delta+S_1+S_2-2}{2}\\
B^2\; u^\frac{-\Delta-S_1-S_2}{2}\\
B^3\; u^\frac{+\Delta+S_1-S_2-2}{2}\\
B^4\; u^\frac{+\Delta-S_1+S_2}{2}
\eea
\right)\;.
\eeq
We note that since $\bP$'s and $\bQ$'s
are not independent due to \eq{QPeq}
there is a nontrivial compatibility condition for their asymptotics \eq{PAM} and \eq{QAM}, which, in particular,
fixes  \cite{Gromov:2014caa}

{
\small
\beqa\label{AA}
A_1A^1&=&\frac{\left(\left(J_1+J_2-J_3-S_2+1\right)^2-\left(\Delta +S_1-1\right)^2\right)
   \left(\left(J_1+J_2-J_3+S_2+1\right)^2-\left(\Delta -S_1+1\right)^2\right)}{-16 i
   \left(J_1+J_2+1\right) \left(J_1-J_3\right) \left(J_2-J_3+1\right)}\\
\nn A_2A^2&=&\frac{\left(\left(J_1-J_2+J_3-S_2-1\right)^2-\left(\Delta +S_1-1\right)^2\right)
   \left(\left(J_1-J_2+J_3+S_2-1\right)^2-\left(\Delta -S_1+1\right)^2\right)}{+16 i
   \left(J_1-J_2-1\right) \left(J_1+J_3\right)\left(J_2-J_3+1\right)}\\
\nn A_3A^3&=&\frac{\left(\left(J_1-J_2-J_3+S_2-1\right)^2-\left(\Delta +S_1-1\right)^2\right)
   \left(\left(J_1-J_2-J_3-S_2-1\right)^2-\left(\Delta -S_1+1\right)^2\right)}{-16 i
   \left(J_1-J_2-1\right) \left(J_1-J_3\right) \left(J_2+J_3+1\right)}\\
\nn A_4A^4&=&\frac{\left(\left(J_1+J_2+J_3-S_2+1\right)^2-\left(\Delta -S_1+1\right)^2\right)
   \left(\left(J_1+J_2+J_3+S_2+1\right)^2-\left(\Delta +S_1-1\right)^2\right)}{+16i
   \left(J_1+J_2+1\right) \left(J_1+J_3\right) \left(J_2+J_3+1\right)}\;.
\eeqa
}
Now we will apply these general formulas, true for any local operator, to our current problem -- the BFKL limit of twist-2 operators.

\section{Quantum Spectral Curve for Twist-2 Operators}

For the twist-2 operators in question, the  charges are fixed to \(J_2=J_3=S_2=0\) and  \(J_1=2\), and we will use the notation \(S_1\equiv S\equiv-1+w\). These operators belong to the so called left-right symmetric sector
for which we have the following reduction  \cite{Gromov:2014caa}:
\begin{eqnarray}
&&\bP^a=\chi^{ac} \bP_c,\qquad  \bQ^i=\chi^{ij} \bQ_j,
\label{LRred}
\end{eqnarray}
where \(\chi\) is the antisymmetric constant \(4\times 4\) matrix with the only nonzero entries
\(\chi^{23}=\chi^{41}=-\chi^{14}=-\chi^{32}=1\).
 From \eqref{LRred}, \eqref{fthorder} and (\ref{tQ=omegaQ}-\ref{omegaQ}) we see  that \(\mu_{23}=\mu_{14},\,\, \omega_{23}=\omega_{14} \),  i.e. we have only $5$ linearly independent components in each of these antisymmetric
 matrices in addition to the non-linear condition of unit Pfaffian.
The asymptotics \eq{PAM} and \eq{QAM} are simplified to
\begin{eqnarray}\label{Plarge}
\bP_a&\simeq&(A_1 u^{-2},A_2 u^{-1},A_3,A_4u)_a\\ \label{Qlarge}
\bQ_j&\simeq&(B_1 u^{\frac{\Delta+1-w}{2}},B_2 u^{\frac{\Delta-3+ w}{2}}, B_3u^{\frac{-\Delta+1-w}{2}}, B_4u^{\frac{-\Delta-3+w}{2 }})_j
\end{eqnarray}
and \eq{AA} reduces to
\begin{eqnarray}
A_1 A_4=-A_1A^1&=&\frac{1}{96i}((5-w)^2-\Delta^2)((1+w)^2-\Delta^2) \label{AB} \\
A_2 A_3=+A_2A^2&=&\frac{1}{32i}((1-w)^2-\Delta^2)((3-w)^2-\Delta^2)\;. 
\label{A3}\end{eqnarray}
Note that one can always make a suitable rescaling to set $A_1=A_2=1$, then $A_3$ and $A_4$
are fixed uniquely by \eq{AB}. This is the normalization we use below in this paper.

\begin{figure}[h]
\center{
\includegraphics[width=.7\linewidth]{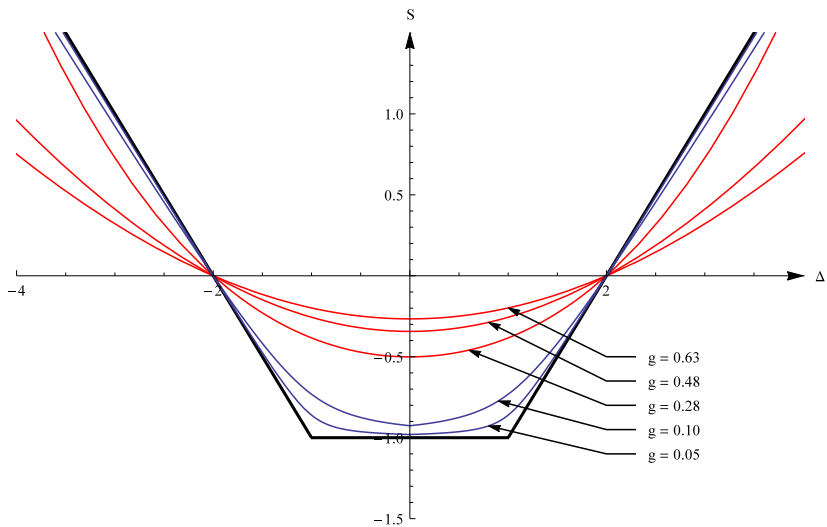}}
\caption{Regge trajectories \(S(\Delta)\) corresponding to the twist 2 operator \({\rm tr}\,Z (D_+)^S Z \) and different values of \(g\).}
\label{Regge_trajectory}
\end{figure}

\paragraph{Prescription for analytic continuation in $S$.}
Before finding the solution for QSC with the above asymptotics, we should precise  the prescription for  analytic continuation in $S$  at the level of QSC. In the section \ref{par_weak_coupling} we explained how the continuation works at weak coupling, in one-loop approximation. We have to translate this prescription into the QSC language. The role of the Q-function \eq{Baxter} in the QSC construction is played by $\mu_{12}$ \cite{Gromov:2013pga}. To make a direct link with the prescription \eq{Qpresc} we consider the $1^{\rm st}$ order equation \eq{fthorder} for $5$ independent components of $\mu_{ab}$. For fixed $\bP_a$ it has $5$ independent solutions whose asymptotics follow from the asymptotics of \(\bP\)'s \eq{Plarge}. One finds that for $\mu_{12}$ one could have one of the following 5 asymptotics \((u^{-1-S},u^{+\Delta-2},u^{-2},u^{-\Delta-2},u^{+S-3})\), where we ordered the possible asymptotics according to their magnitude in the BFKL regime, i.e. when  $S\to -1$ and $0<\Delta<1$
(see Fig.\ref{Regge_trajectory}).
Note that for the usual perturbative regime considered in the introduction we have $\Delta\sim S+2$
and thus we can recognize in the first two solutions $Q_2$ and $Q_1$ correspondingly!
This motivates the prediction which was put forward and tested by \cite{Gromov:2014bva}, stating that
 in order to analytically continue the QSC to non-physical domain of non-integer $S$
one should relax the power-like behavior of $\mu_{ab}$ (required for all physical states) allowing for
the following leading and subleading terms in the asymptotics:\beq\label{mu12leading}
\mu_{12}\sim {\rm const}\; u^{+\Delta-2}+e^{2\pi u}{\rm const}\; u^{-1-S}+\dots.
\eeq
This is the generalization of \eq{Qpresc} to a finite coupling.
In \cite{Gromov:2014bva} it was proven that with this prescription
there is a unique solution for any coupling, at least in the vicinity of $S=0$ up to the order $S^2$ inclusively.
We also know that at weak coupling there exists the unique solution for these asymptotics.
We consider this to be a strong indication towards uniqueness of such solution to an arbitrary number of loops,
which would be however very interesting to prove rigorously.
As we will also see below, such a solution is also unique   in the BFKL regime.
As this asymptotic is also consistent with the asymptotics for the physical states it should thus provide
an analytic continuation of the physical solutions to an arbitrary non-integer $S$.

\subsection{Leading Order solution for $\bP\mu$-system}
The logic of this section is the following:
we begin by arguing a certain scaling in the small parameter $w\equiv S+1$ for various quantities
and then write an ansatz for $\bP_a$ and $\mu_{ab}$.
First, we assume that $\bP_a\sim w^0$, in accordance with its large $u$ asymptotics \eqref{AB}.
Second, we keep in mind that the BFKL regime is still a regime of weak coupling, even though it re-sums
all singularities of $\Delta$ of the type $(g^2/w)^n$.
This means that all cuts are collapsed to a point and, in a generic situation,
all functions should be regular on the entire complex plain. However, there could be some special cases where this
rule is violated. Namely, consider a function $f(u)=\frac{1}{(g x)^2}$
where  $x=\frac{u+\sqrt{u^2-4g^2}}{2g}$ is the Zhukovsky variable. In the BFKL regime $g\to 0$
\beq\la{fft}
f(u)\simeq \frac{1}{u^2}\;\;,\;\;\tilde f(u)=\frac{x^2}{g^2}\simeq \frac{u^2}{g^4}\simeq \frac{u^2}{w^2\Lambda^2}\;.
\eeq
This shows that in principle even in the limit when the cut totally disappears some
functions still can develop a singularity by the cost of being very large (but regular) on the next sheet.
Note, that this exception is clearly not applicable to $\mu$
since $\tilde \mu$ is the same as shifted \(\mu(u+i)\), so that  both
$\mu$ and $\tilde\mu$ should be of the same order and thus are regular at the leading order in $w$.

Nevertheless, $\bP_1(u)$ must be exactly a function of this type.
Indeed, at large $u$  it behaves as $1/u^2$ and has no other singularities except for the cut.
From that we conclude that we must have a double pole at zero or even a stronger singularity.
The residue at the double pole is uniquely fixed at this order by \eq{Plarge}, i.e. $\bP_1(u)=1/u^2$
whereas we will see that the
stronger singularities could indeed appear at the next order in $w$.
This $1/u^2$ singularity at zero
implies that at the next sheet the function scales with our expansion parameter as
$\tilde \bP\sim 1/w^2$, which is only possible if at least some components of $\mu_{ab}$
scale  as $\mu_{ab}\sim 1/w^2$, as we can see from
\eq{Pt}. Consequently, $1/w^2$ will propagate via \eq{Pt} into all components of $\bP_a$.
To summarize, we have to find a solution with the following scaling in $w$:
\beq
\bP_a\sim w^0\;\;,\;\;\mu_{ab}\sim w^{-2}\;\;,\;\;\tilde\bP_a\sim w^{-2}\;.
\eeq

This scaling will lead us to an ansatz which
we then plug into the ${\bf P}\mu$-system to fix the remaining freedom.
We start from the \(\bP\)-functions which have the simplest analytic structure: only one short cut on the main sheet, and   integer powers in asymptotics. We can thus uniformize them by Zhukovsky
map \(u=g(x+1/x)\), with the inverse $x(u)$ introduced above, such that \(\tilde x=1/x\),
and expand \(\bP_a\) into the Laurent series around $x=0$ \cite{Marboe:2014gma}
\begin{eqnarray}\label{Pexp}
\bP_a=\sum_{n=-1}^{\infty}\frac{c_{a,n}}{x^{n}}.
\end{eqnarray}
It is guaranteed to converge for $|x|>|1/x(2g+i)|$
which allows to cover the whole upper sheet and even a finite part of the next sheet
and leads to the corresponding ansatz for $\tilde\bP_a$:
\beq\la{pta}
\tilde\bP_a=\sum_{n=-1}^{\infty}c_{a,n}x^{n}\;.
\eeq
To reduce the number of coefficients we note that for our observable there must be a parity symmetry $u\to -u$ (or equivalently $x\to -x$).
Of course all $\bP\mu-$system equations are invariant w.r.t. this transformation,
which means that this symmetry in general maps one solution to another. As we know that our state with these quantum numbers is unique
we conclude  of course that the parity transformation should map our solution to an equivalent solution.
Using the arguments similar to \cite{Gromov:2014caa} in Sec.4.4.2
it is posible to show that we can fix the remaining freedom in the construction and choose solution where
$\bP_a$ are mapped to themselves. From the asymptotics we see that $\bP_1$ is even, whereas $\bP_2$ is odd etc. To summarize, we impose
\beq
\bP_{1},\bP_{3}\;\;-\;\;{\rm even}\;\;\;\;,\;\;\;\;
\bP_{2},\bP_{4}\;\;-\;\;{\rm odd}\;.
\eeq
Similarly $\mu$'s should be covariant under the parity transformation.
As the parity transformation is sensitive to the choice of cuts we should also take $\hat \mu_{ab}$ with short cuts
to be covariant under the parity transformation which implies that $\hat\mu_{12}$ is even, however as a consequence of this $\check\mu_{12}$
should transforms nontrivially
$
\check\mu_{12}(-u)=\tilde{\check\mu}_{12}(u)={\check\mu}_{12}(u+i)
$
which by itself implies, by changing $u\to u-i/2$, that $\mu_{12}^+$ is even. To summarize, we have
\beq\la{evenodd}
\mu_{12}^+,\;\mu_{14}^+,\;\mu_{34}^+\;\;-\;\;{\rm even}\;\;\;\;,\;\;\;\;
\mu_{13}^+,\;\mu_{24}^+\;\;-\;\;{\rm odd}\;.
\eeq

This conditions allows us to drop each other coefficient in our ansatz
\(c_{2,2n}=c_{1,2n-1}=c_{4,2n}=c_{3,2n-1} \equiv 0\) at any \(n\).
It also follows from \eqref{Plarge} that the coefficients \(c_{2,-1}=c_{1,0} \equiv 0 \).
After that we still have one-parametric scalar freedom in our construction
\cite{Gromov:2014bva}:   \begin{eqnarray}\label{H-tr}
&&\bP_3\to\bP_3  +\gamma \bP_1\,,\qquad \bP_4\to\bP_4  -\gamma \bP_2\,,\notag\\
&& \mu_{14}\to \mu_{14}-\gamma\mu_{12}\,,\qquad \mu_{34}\to \mu_{34}+2\gamma\mu_{14}-\gamma^2 \mu_{12}\,,
\end{eqnarray}
preserving the leading \( u\to\infty\) asymptotics but modifying the subleading ones.
This allows to fix in addition \(c_{3,2}=0\). Finally, as we use the normalization with $A_1=A_2=1$
we have to fix \(c_{2,1}=\frac{1}{\sqrt{\Lambda w}}\) and \(c_{1,2}=\frac{1}{\Lambda w}\).

Let us now restrict the possible scaling of the coefficients $c_{a,n}$ in the BFKL limit.
In this limit \(g^2\sim w\to 0\) the Zhukovsky cut shrinks into a point and the $x(u)$ becomes
\beq
x(u)=\frac{u}{g}-\frac{g}{u}-\frac{g^3}{u^3}+\dots\;.
\eeq
To satisfy the scaling \(\tilde \bP_a\sim 1/w^2\sim 1/g^4\) the coefficients \(c_{a,n}\) should become smaller and smaller with \(n\) and in general they scale as \(c_{a,n}\sim g^{n-4}\). We thus denote
\beq\la{scalingc}
c_{m,n}=g^{n-4}\sum_{k=0}^{+\infty}c_{m,n}^{(k)}w^k.
\eeq
where \(c_{m,n}^{(k)}\) are already \(\sim 1\). To the leading order in $w$ we thus simply get
\begin{align}
& \bP_1\simeq\frac{1}{u^2}\;\;,\;\;\bP_2\simeq\frac{1}{u}\;\;,\;\; \bP_3\simeq A_3^{(0)}
\;\;,\;\; \label{PLO}
 \bP_4\simeq A_4^{(0)} u+\frac{c_{4,1}^{(1)}}{\Lambda u}\;
.
\end{align}
where from \eqref{AB} we have \(A_4^{(0)}=-\frac{i}{96}(\Delta^2-1)(\Delta^2-25)\), \(A_3^{(0)}=-\frac{i}{32}(\Delta^2-1)(\Delta^2-9)\) and the only coefficient to fix is \(c_{4,1}^{(1)}\).
Now when $\bP$'s are essentially fixed, we can use the $5^{th}$ order equation \eq{fthorder}
to find $\mu$. Note that $\bP_1$ and $\bP_2$ are singular whereas $\tilde \bP_a$ should be regular at $u=0$.
This is only possible if $\mu_{ab}$ are regular and have a sufficient amount of zeros at $u=0$.
This observation  singles out one solution out of $5$ possible ones, with $\mu_{12}\sim u^{-S-1}$ for which
all components of $\mu_{ab}$ have a polynomial asymptotic for large $u$: \((\mu_{12},\mu_{13},\mu_{14},\mu_{24},\mu_{34})_{LO}\sim(u^0,u^1,u^2,u^3,u^4)\).
To find this solution we plug a polynomial ansatz into \eq{fthorder}, to find
\beqa
&& \mu_{12}^+\simeq+\cal P, \\
&& \mu_{13}^+\simeq-\frac{{\cal P}}{16} i u  \left(\Delta ^2-1\right)^2, \\
&& \mu_{14}^+\simeq-\frac{{\cal P}}{128} i \left(4 u^2+1\right)  \left(\Delta ^2-1\right)^2, \\
&& \mu_{24}^+\simeq-\frac{{\cal P}}{192} i u\left(4 u^2+1\right) \left(\Delta^2-1\right)^2, \\
&& \mu_{34}^+\simeq-\frac{{\cal P}}{49152}\left(16 u^4-8 u^2-3\right) \left(\Delta ^2-1\right)^4,
\eeqa
and in addition \eq{fthorder} also to fix \(c_{4,1}^{(1)}=-\frac{i(\Delta^2-1)^2\Lambda}{96} \).
Thus we fix $\mu_{ab}$ up to  a common multiplier ${\cal P}$.
As we deal with a finite difference equation  this multiplier could be only an i-periodic function,
which has to be chosen in accordance with the prescription \eq{mu12leading}
and which respects the parity \eq{evenodd}. The most general choice is
\beq
\CP=C_1+C_2\cosh^2(\pi u)
\eeq
for some constants $C_1$ and $C_2$.
Thus we have only two constants to fix and still several nontrivial conditions to satisfy, namely
\eq{muPPt-eq} and \eq{Pt}. From \eq{Pt} we find
\small
\beq\la{bp1}
{\tilde \bP}_1 \simeq -\frac{i \left(\Delta ^2-1\right) {\cal P}^-}{4}
,\;\;
{\tilde \bP}_2 \simeq -\frac{i \left(\Delta ^2-1\right) u
   {\cal P}^-}{4}
,\;\;
{\tilde \bP}_3 \simeq -\frac{\left(\Delta ^2-1\right)^3 u^2 {\cal P}^-}{128}
,\;\;
{\tilde \bP}_4  \simeq -\frac{\left(\Delta
   ^2-1\right)^3 \left(u^3+u\right) {\cal P}^-}{384}
\eeq
\normalsize
To fix $C_1$ and $C_2$ we note that from the ansatz for $\tilde\bP_1$
\eq{pta} we should have
\beq
\tilde \bP_1= \frac{u^2+{\cal O}(u^4)}{w^2\Lambda^2}+{\cal O}\left(\frac{1}{w}\right)
\eeq
which is also clear from our basic discussion \eq{fft}. Comparing with \eq{bp1} we fix
\beq
C_1=0\;\;,\;\;C_2=\frac{4 i}{\pi^2\Lambda^2 w^2 (\Delta^2-1)}\;.
\eeq
We found a consistent solution with no free parameters left. We also might expect that we could  get a relation between the energy \(\Delta\) and the coupling \(\Lambda\), which are so far completely independent. However, it is not the case at this order. The reason for this is that we are not able to use efficiently the remaining condition \({\rm Pf}\mu_{ab}=1\) because the l.h.s. is of order \(1/w^4\) and with our precision we cannot distinguish the \(1\) in the r.h.s. from any other finite number. We would have to continue the procedure to the next \(4\) orders in \(w\) until we get this condition to work. As we will see, a much more efficient way to overcome this difficulty is to pass to \(\bQ\omega\)-system.

\subsection{Next-to-Leading-Order solution}
We can also extend the consideration of the previous section to the next order in $w$.
Using the ansatz \eq{Pexp} and \eq{scalingc} we get, up to $w^1$ terms
\begin{align}
& \bP_1\simeq\frac{1}{u^2}+\frac{2\Lambda w}{u^4}, \notag \\
& \bP_2\simeq\frac{1}{u}+\frac{2\Lambda w}{u^3}, \notag\\
& \bP_3\simeq A_3^{(0)}+A_3^{(1)}w, \notag \\
& \bP_4\simeq A_4^{(0)} u-\frac{i(\Delta^2-1)^2}{96 u}+\left(A_4^{(1)}u+\frac{c_{4,1}^{(2)}}{u\Lambda}-\frac{i(\Delta^2-1)^2 \Lambda}{48u^3} \right)w. \label{Pmain}\end{align}
where from \eqref{AB} \(A_3^{(1)}=-\frac{i}{4}(\Delta^2-3)\), \(A_4^{(1)}=-\frac{i}{12}(\Delta^2+5)\).
Again there is only one missing constant \(c_{4,1}^{(2)}\).
To fix it we have to proceed further to find \(\mu\) in the NLO.
At this order the solution cannot be just a polynomial as the asymptotic $u^{-w}$ of $\mu_{12}$ suggests
that the ansatz is more complicated. We discuss details of this calculation in Appendix \ref{NLOP}
where we find that the missing constant is
\begin{equation}\label{c312}
c_{4,1}^{(2)}=-\frac{i\Lambda}{24}(\Delta^2-1)\left[2(\Delta^2-1)\Lambda-1\right].
\end{equation}
We will use this result to
finding the NLO for \(\bQ\)-functions and also the LO result for the BFKL  dimension.

\subsection{Passing to $\bQ\omega$-system}

An important step in our calculation is to switch now to the $\bQ\omega$-system.
It is especially easy having at hand the eq.\eqref{QPeq}.
We simply plug the already known $\bP$'s \eqref{PLO} into \eqref{QPeq} an get a 4-th order linear finite difference equation on \(\bQ_j\) with explicit polynomial coefficients.
As a good sign that we are on the right track, the finite difference operator of this 4-th order equation  can be nicely factorized as follows:
\begin{eqnarray}\label{4thOrderBFKL}
&&\left[(u+2i)^2D+(u-2i)^2D^{-1}-2u^2-\frac{17-\Delta^2}{4}\right]\left[D+D^{-1}-2-\frac{1-\Delta^2}{4u^2}\right]{\bf Q}=0
\end{eqnarray}
where \(D=e^{i\partial_u}\) is the shift operator.
This implies that two out of four \(\bQ\)-functions satisfy
the 2-nd order equation
\begin{eqnarray}\label{BaxterLO0}
\bQ_j \frac{\Delta ^2-1-8 u^2}{4 u^2}
+\bQ_j^{--} +\bQ_j^{++}
   =0.
\end{eqnarray}
Even before solving this equation, it is easy to check that there are two independent solutions with the large $u$
asymptotics $u^{\frac{+\Delta+1}{2}}$ and $u^{\frac{-\Delta+1}{2}}$,  which indictes,  together with \eq{Qlarge},
that they can be identified as $\bQ_1$ and $\bQ_3$.

Notice that (\ref{BaxterLO0}) after redefinition
\(Q=\frac{\bQ_j}{u^2}\) is precisely the famous \(sl(2,\mathbb{C})\) Baxter equation defining, through Sklyanins
separation of variables method, the Pomeron LO BFKL wave function \cite{Lipatov:1993yb,Faddeev:1994zg,DeVega:2001pu,Derkachov:2002pb,Derkachov:2002wz,Derkachov:2001yn}! It can be solved, taking into account the asymptotics and the  UHP analyticity, in terms of a hypergeometric function
\begin{eqnarray}\la{Q1Q3}
&&{\bf Q}_1=2 i u \, _3F_2\left(i u+1,\frac{1}{2}-\frac{\Delta }{2},\frac{\Delta }{2}+\frac{1}{2};1,2;1\right)\\
&&{\bf Q}_3(u)={\bf Q}_1(-u) \sec \frac{\pi\Delta}{2}+ {\bf Q}_1(u)\left[-i\coth(\pi  u)
+ \tan \frac{\pi\Delta}{2}\right].   \label{Q1,3}\notag
\end{eqnarray}
where ${\bf Q}_1$ is a solution chosen to be analytic in the upper half plane,
and it can be read off from the papers \cite{Lipatov:1993yb,Faddeev:1994zg,DeVega:2001pu,Derkachov:2002pb,Derkachov:2002wz,Derkachov:2001yn}.
At the same time, it serves as a building block for the second solution $\bQ_3$,
which has a smaller (for $\Delta>0$) asymptotic. To build $\bQ_3$
we use that ${\bf Q}_1(-u)$ is also an independent solution of the same
Baxter equation, which, however, has poles in the upper half plane.
We cancel these poles by adding ${\bf Q}_1(u)$ we a suitable periodic coefficient,
and finally extract ${\rm const}\times {\bf Q}_1(u)$ to ensure the right large $u$ asymptotics.
The choice appears to be unique.

Although  for the rest of the paper  we will not need to determine  the other two \(\bQ\)-functions,\(\bQ_2,\bQ_4\)  we give here for completeness   the inhomogeneous equation following from \eqref{4thOrderBFKL}  expressing them
through  \(\bQ_1,\bQ_3\)
\begin{equation}
\bQ_{j}^{++}+\bQ_{j}^{--}-\left(2+\frac{1-\Delta^2}{4u^2} \right)\bQ_{j}
\,=\,\frac{\pi^2(\Delta^2-1)^2}{16\cos\frac{\pi\Delta}{2}}
\left(\frac{\bQ_{j-1}^{++}}{(u+i)^2}+\frac{\bQ_{j-1}^{--}}{(u-i)^2}-2\frac{\bQ_{j-1}}{u^2}\right),
\qquad j=2,4.  \end{equation}
The \(\bQ_2,\bQ_4\) may become useful for the calculation of dimension in NLO and NNLO of the BFKL approximation.

\paragraph{NLO Baxter equation.}
Similarly we can use our knowlege of the NLO ${\bf P}'s$
\eq{Pmain} to construct the NLO Baxter equation which takes into account the \({\cal O}(w)\) terms in \eqref{QPeq}.
For that we have to plug there   \(\bP_{a}\)   given by \eqref{Pmain}, \eqref{c312}. Again, this 4-th order finite difference equation appears to be factorizable and  as a  result we get the following 2-nd order equation
for $\bQ_1$ and $\bQ_3$, generalizing \eq{BaxterLO0}:
\begin{eqnarray}\label{Baxter-w}
\bQ_j \left(\frac{\Delta ^2-1-8 u^2}{4 u^2}+w \frac{\left(\Delta ^2-1\right) \Lambda-u^2 }{2 u^4}\right)+\bQ_j^{--} \left(1-\frac{i w/2}{u-i}\right)+\bQ_j^{++}
   \left(1+\frac{i w/2 }{u+i}\right)=0\,,\qquad j=1,3.\notag\\\end{eqnarray}
Its solution can be found using Mellin transformation method of \cite{Faddeev:1994zg} and is given in Appendix~\ref{AppQ}.
For our present goal -- the calculation of BFKL dimension, we only need a simple fact about the NLO $\bQ$.
Namely, we want to know its behavior around $u=0$. This information is easier to extract
directly from the Baxter equation \eq{Baxter-w} by shifting $u\to u+i$.
Recalling  that $\bQ_j$ must be regular in the upper-half-plane we obtain from the second term
\beq
\left(1-\frac{iw}{2u}\right)\bQ_j(u)={\rm regular\;at\;}u\sim0
\eeq
which gives the relation between the behaviour at the origin of the leading in $w$  order $\bQ_j^{(0)}$ and the
subleading order $\bQ_j^{(1)}$:
\beq\la{Q1oQ0}
\frac{\bQ_j^{(1)}(u)}{\bQ_j^{(0)}(u)}= +\frac{iw}{2u}+{\cal O}(u^0)\;\;,\;\;j=1,3\;.
\eeq
The strategy is now to compute this ratio independently, using the $\bQ\omega$-system. Matching these two results we
will recover  the BFKL Pomeron eigenvalue.

\subsubsection{Going to the next sheet}
So far we mostly recycled the information from $\bP\mu$-system into $\bQ$'s.
To get something new we have to work a bit harder and reconstruct $\omega$'s. This will allow us
to compute, for example, $\tilde \bQ_3(0)$ from which we will instantly determine the pole in the ratio
\eq{Q1oQ0}. Let us remind the relation between $\mu_{ab}$ and $\omega_{ij}$.
They are related to each other by a Q-function with $2+2$ indices
\beq\la{mo}
\mu^{ab}=2{\cal Q}^{a|i-}{\cal Q}^{b|j-}\;\omega_{ij}
\eeq
where ${\cal Q}_{ab|ij}$ can be decomposed in terms of $\bP_a$ and $\bQ_i$
and as a result should be $\sim w^0$. This implies that at least one of the components of $\omega_{ij}$
should scale as $1/w^2$. As we know,  at the leading order in $w$, up to a periodic function,
$\mu_{12}\sim u^{-S-1}$ and we can precisely identify it with $\omega_{24}$ which should be
$\sim 1/w^2$ whereas other components should be smaller.
We will see that the consistent scaling is
 \(\omega_{12} \sim \omega_{14} \sim \omega_{34} \sim w^0\) and \(\omega_{24} \sim w^{-2}\) and \(\omega_{13} \sim w^2\).
 The reason why $\omega_{13}$ appear to be $w^2$ is due to the fact that
  this component multiplies $\omega_{24}$
 in the Pfaffian which is set to $1$.

With this insight coming from $\mu$'s we can see that only two terms survive in
the relation for $\tilde \bQ_1$ and $\tilde \bQ_3$ since the term with $\omega_{13}$ are too small
\begin{align}\la{bQt}
& \tilde{\bQ}_1(u)=+\omega_{14}(u)\bQ_1(u)+\omega_{12}(u)\bQ_3(u), \\
& \tilde{\bQ}_3(u)=+\omega_{34}(u)\bQ_1(u)-\omega_{14}(u)\bQ_3(u).\nn
\end{align}
Note that since these components are suppressed compare to $\omega_{24}$ no explicit information about
their form can be extracted from $\bP\mu$-system at the given order in $w$. At the same time we can say that
$\omega_{24}= B \sinh^2(\pi u)$ just because the Q-function in \eq{mo} has a power-like asymptotics and all the exponents
can only originate from the factor $\omega_{24}$.
Another thing to notice, is that ${\bQ}_1,{\bQ}_3$ decouple from the rest of the $\bQ\omega$-system.
This explains  in particular the mysterious factorization of the 4th order equation \eq{4thOrderBFKL}.

We will now fix $\omega$'s appearing in \eq{bQt}
using some elementary properties of $\bQ_1$ and $\bQ_3$
found explicitly in \eq{Q1Q3}. We already pointed out that $\bQ_1(-u)$ and $\bQ_3(-u)$
would also be solutions of the same finite-difference equation and thus they can be re-expanded
in terms of the basis $\bQ_1(u),\;\bQ_3(u)$. Right from
\eqref{Q1Q3} we have
\begin{align}\la{Qmin}
& \bQ_1(-u)=+\bQ_1(u)\frac{i\cosh\left(\pi\left(u+\frac{i\Delta}{2} \right) \right)}{\sinh(\pi u)}+\bQ_3(u)\cos\left(\frac{\pi\Delta}{2} \right), \\
& \bQ_3(-u)=-\bQ_1(u)\frac{\cos\left(\frac{\pi\Delta}{2}\right)}{\sinh^2(\pi u)}+\bQ_3(u)\frac{i\cosh\left(\pi\left(u-\frac{i\Delta}{2} \right) \right)}{\sinh(\pi u)}.
\end{align}
this equation in many respects is similar to the equation we want to recover, \eq{bQt}.
Indeed, both $\tilde \bQ(u)$ and $\bQ(-u)$ are analytic below the real axis, and the
coefficients in the r.h.s. are periodic functions of $u$ as $\omega$'s should be.
We have to find a relation between $\tilde\bQ(u)$ and $\bQ(-u)$ which we may already expect to be simple.
Combining \eq{Qmin} and \eq{bQt} we can write
\begin{align}\label{tilde_system}
& \tilde{\bQ}_1(u)=a_{11} \bQ_1(-u)+a_{13} \bQ_3(-u), \\
& \tilde{\bQ}_3(u)=a_{31} \bQ_1(-u)+a_{33} \bQ_3(-u).
\end{align}
{\it A priori} $a_{ij}$ are some periodic functions of $u$. Let us show that they must be constants.
Firstly, they should have no poles. That is because both $\tilde\bQ(u)$ and $\bQ(-u)$
could not have any poles below the real axis, from the explicit form of $\bQ_1$ and $\bQ_3$
we can verify they do not vanish at $u=-i n$ and cannot cancel the poles themselves,
furthermore the cancelation of the poles
between the two terms
in the r.h.s. is impossible as $\bQ_3$ decays faster and soon becomes negligible comparing to the term with $\bQ_1$
when we go down in the complex plane. Secondly,
$a$'s cannot grow exponentially at infinity as $\bQ(-u)$'s and $\tilde\bQ(u)$'s behave power-like in the lower-half plane.
Therefore, according to the Liouville theorem these coefficients are constants.
Thus our problem of finding $\omega$'s
is already simplified enormously and reduced to the problem of finding a few constants.

Next, we have to remember that $\tilde\bQ_i$ is an analytic continuation of $\bQ_i$ and so they should match at $u=0$
\beq
\tilde\bQ_1(0)=\bQ_1(0)=0\;\;,\;\;\tilde\bQ_3(0)=\bQ_3(0)\neq 0
\eeq
which fixes $a_{13}=0$ and $a_{33}=1$.
Now we can combine this information with \eq{Qmin}
to see what it implies for $\omega$'s.
We notice that in \eq{bQt} there are only $3$ different coefficients,
which gives a nontrivial constraint on $a$'s
\beq
i a_{11}\frac{\cosh\left(\pi u+\frac{i\pi\Delta}{2}\right)}{\sinh(\pi u)}=-a_{31}\cos\frac{\pi\Delta}{2}
-i\frac{\cosh\left(\pi u-\frac{i\pi\Delta}{2}\right)}{\sinh(\pi u)}
.\eeq
On the first sight, it seems to be impossible to satisfy with constant $a_{11}$ and $a_{31}$ for any $u$. Luckily,
it is solved at once for $a_{11}=-1$ and $a_{31}=-2\tan\frac{\pi\Delta}{2}$!
From where we obtain in particular
\begin{align}
\nn & \tilde{\bQ}_1(u)=-\bQ_1(-u), \\
\la{Qt3}& \tilde{\bQ}_3(u)=+\bQ_3(-u)-2\tan\left(\frac{\pi\Delta}{2} \right)\bQ_1(-u).
\end{align}
One can also read $\omega$'s from this expression, which is done in Appendix \ref{app_omega}.

\subsection{LO BFKL dimension}

We are just one step away from the main result -- BFKL dimension.
For that we notice that the knowledge of, say, $\bQ_3$ and $\tilde\bQ_3$
in the $u\sim 1$ scaling gives an access to the leading singularity in $\bQ_3$ at $u=0$
to all orders in $w$. Indeed, the combination
$\bQ_3-\tilde\bQ_3$
changes the sign when we go under the Zhukovsky cut
and thus is proportional to $\sqrt{u^2-4g^2}$. In other words $\frac{\bQ_3-\tilde\bQ_3}{\sqrt{u^2-4g^2}}$
does not have the cut $[-2g,2g]$ anymore and thus it is regular in  $g\ll u\sim 1$ scaling. Same is true about
the even combination $\bQ_3+\tilde\bQ_3$. Thus we can rewrite
\beq
\bQ_3=
\frac{\bQ_3-\tilde\bQ_3}{2\sqrt{u^2-4g^2}}\sqrt{u^2-4g^2}
+
\frac{\bQ_3+\tilde\bQ_3}{2}
=
\left[\frac{\bQ_3-\tilde\bQ_3}{\sqrt{u^2-4g^2}}\right]\left(-\frac{\Lambda w}{u}-\frac{\Lambda^2 w^2}{u^3}+\dots\right)+{\rm regular}
\eeq
note that we know explicitly the expression
in the square brackets at the leading order in $w$ from \eq{Qt3} and \eq{Q1Q3}. Its small $u$ expansion gives
\beq
\frac{\bQ_3-\tilde\bQ_3}{\sqrt{u^2-4g^2}}=
2i\bQ_3(0)\Psi(\Delta)+{\cal O}(\omega)+{\cal O}(u)
\eeq
where
\beq
\Psi(\Delta)\equiv \psi\left(\frac{1}{2}-\frac{\Delta}{2}\right)+\psi\left(\frac{1}{2}+\frac{\Delta}{2}\right)-2\psi(1)\;.
\eeq
From that we can immediately find that the pole in $u$ at the first order in $w$ should be
\beq \label{PoleQ3}
\bQ_3(u)=-\frac{2i\bQ_3(0)\Psi(\Delta)\Lambda w}{u}+{\rm regular}+{\cal O}(w^2).
\eeq
this compared with \eq{Q1oQ0} lead to
\beq
-4\Psi(\Delta)\Lambda=1\;.
\eeq
This is precisely the formula  \eq{KL}
for the eigenvalue
of the QCD BFKL kernel,
or, equivalently
for the dimension of  twist-2 operator in BFKL approximation at the leading Regge singularity!

\section{Discussion}

In this paper we managed to  reproduce the dimension of twist-2 operator of  ${\cal N}=4$ SYM theory in the 't~Hooft limit in the leading order (LO) of the  BFKL regime directly from  exact equations for the spectrum of local operators called the Quantum Spectral Curve -- QSC. This result  is a very non-trivial confirmation of  the general validity of this QCS approach and of the whole program of integrability of the spectral problem in AdS/CFT --    S-matrix and asymptotic Bethe ansatz, TBA, Y-system, FiNLIE equations, etc. In particular, this is one of a very few examples of all-loop calculations, with all wrapping corrections included, where the integrability result can be checked by direct Feynman graph summation of the original BFKL approach.  An obvious step to do in this direction is to compute the NLO correction to the twist-2  dimension
 from QSC and compare to the direct BFKL computation of   \cite{Kotikov:2002ab}. Many of the elements of the NLO construction, such as the NLO Baxter equation for \(\bQ\)-functions, are present already in this paper,
 but the most difficult ingredient -- the formula of the type \eqref{PoleQ3} for the leading singularity, has yet to be derived. Of course, the ultimate goal of the BFKL approximation to QSC would be to find an algorithmic way of generation of any BFKL correction (NNLO, NNNLO, etc) on Mathematica program, similarly to the one for the weak coupling expansion via QSC, proposed by \cite{Marboe:2014gma}. It would be also very interesting to build  numerically from the QSC  the twist-2 dimension as a continuous function of spin \(S\in \mathbb{R}\) qualitatively described in \cite{Brower:2006ea}.  We also hope that our approach will allow to understand deeper the similarities and differences of N=4 SYM and the pure Yang-Mills theory (multicolor QCD)  starting from the BFKL approximation, regarding the well known fact that, at least in the 't~Hooft limit, N=4 SYM Feynman graphs capture an important part of all QCD graphs and in the  LO BFKL the results simply coincide.

We also hope that the methods of QSC presented here will be inspiring for construction of the systematic strong coupling expansion in N=4 SYM.
A deeper insight into the structure of QSC will be needed to approach the whole circle of these complex problems.

\begin{acknowledgments}

\label{sec:acknowledgments}
We thank     S.~Caron-Huot, R.~Janik, G.~Korchemsky,  L.~Lipatov, F.~Levkovich-Maslyuk, G. Sizov, E.~Sobko and S.~Valatka for discussions. The research
of N.G., V.K., leading to these results has received funding from the People Programme
(Marie Curie Actions) of the European Union's Seventh Framework Programme FP7/2007-2013/ under
REA Grant Agreement No 317089.  The research of V.K. leading to these results has received funding from the European Research Council (Programme "Ideas" ERC-2012-AdG 320769 "AdS-CFT-solvable"), from the ANR
grant StrongInt (BLANC- SIMI- 4-2011) and from the ESF grant HOLOGRAV-09- RNP- 092. The research of M.A. leading to these results has received funding from the People Programme (Marie Curie Actions) of the European Union's Seventh Framework Programme under Grant Agreement No 317089, from the Russian Scientific Foundation under the grant 14-12-01383 and from the 2013 Dynasty Foundation Grant. M.A. and V.K.  are  very grateful to the  Institute
for Advanced Study (Princeton), where a part of this work was done,  for hospitality. V.K. also
thanks the Ambrose Monell Foundation, for the generous support.
N.G. also thanks SFTC for support
from Consolidated grant number ST/J002798/1.
N.G. acknowledges support
of Russian Science Foundation, grant number 14-22-00281.
\end{acknowledgments}

\appendix

\section{Derivation of the 4-th order equation \eqref{QPeq} for $\bf{Q}$}
\label{Der4orderEq}
Following the discussion in \cite{Gromov:2014caa}
we begin by picking a few QQ-relations out of many possible:\begin{eqnarray}\label{QPQ}
{\cal Q}_{a|j}^+ - {\cal Q}_{a|j}^-=\bP_a\bQ_j\;,\qquad
\bQ_j\equiv-\bP^a {\cal Q}^\pm_{a|j}\;.
\end{eqnarray}
The first one follows directly from \eqref{QQbf} with \(A=I=\emptyset\). The second one  is also an algebraic consequence of general QQ-relations \eqref{QQbb}, \eqref{QQff} and \eqref{QQbf}.
It is shown in   \cite{Gromov:2014caa} (see eqs(4.5)-(4.7) therein) that all Q-functions can be obtained in a simple way through 3 ones -- \(\bP_a,\, \bQ_j\;\) and \({\cal Q}_{a|j}\). Thus, expressing \(\bP^a\equiv{\cal Q}_{\bar a|1234}\), where \(\bar a=\{1234\}\backslash a\) is a subset complimentary to \(a\), through these 3 types of  Q-functions we can prove the second of eqs.\eqref{QPQ}.

     We see that the function ${\cal Q}_{a|j}$ is  designed to ``rotate" $\bP$ into $\bQ$.
The strategy is to exclude ${\cal Q}_{a|j}$ and related $\bP$ and $\bQ$ directly.
Shifting the argument \(u\) in the second of these two equations   by \(\pm i,\pm 2i\) and then using there the first equation to bring all shifted arguments in  \({\cal Q}_{a|j}\) to the the same one we obtain a linear system of 4 equations     \begin{eqnarray}
\bP^{a[-3]} {\cal Q}_{a|j}&=&\bQ_j^{[-3]} -\bQ_j^{[-1]} (\bP^{a[-3]}\bP_a^{[-1]})\notag\\
\bP^{a[-1]} {\cal Q}_{a|j}&=&\bQ_j^{[-1]} \notag\\
\bP^{a[+1]} {\cal Q}_{a|j}&=&\bQ_j^{[+1]} \notag\\
\bP^{a[+3]} {\cal Q}_{a|j}&=&\bQ_j^{[+3]} -\bQ_j^{[+1]} (\bP^{a[+3]} \bP_a^{[+1]})
\label{LinSys}\end{eqnarray}
from which we can express \({\cal Q}_{a|j}\) in terms of \(\bP_a,\,\bP^a\) and \(\bQ_j\).
Now, taking  the second of two equations \eqref{QPQ} in the shifted form \(\bQ_j^{[5]}\equiv-\bP^{a[5]} {\cal Q}^{[4]}_{a|j}\;\)
and using there again   the first equation to bring all shifted arguments in  \({\cal Q}_{a|j}\)
to the the same one, we get, together with \eqref{LinSys}, a system of 5 linear equations on only 4 functions   \(\gQ_{a|j}\) (the second index is fixed in all equations).        From their compatibility we obtain the closed 4-th order linear equation     \eqref{QPeq} with coefficients expressed only through  \(\bP_a,\,\bP^a\).

\section{Computation of $\bP_j$ in the NLO }
\label{NLOP}

The asymptotics of \(\mu\)'s in this order contains a logarithmic correction.
For example, from \eqref{mu12leading} it follows that, up to the exponential factor,  \(\mu_{12}\sim u^{-S-1}\sim 1-w\log u\).
This shows that $\mu_{ab}$ can no longer be a polynomial times exponentials.
Nevertheless, we show here that the non-polynomial part can be identified easily and we will have to find a few coefficients in the polynomial part
as we did for the leading order.

As we discuss in the main text, in the expression \eq{mo}, relating $\mu_{ab}$ to $\omega_{ij}$
only the term with $\omega_{24}$ (or equivalently $\omega^{13}$) survives since the other terms are suppressed by $w^2$.
Therefore we can write \eqref{muomega} as
\begin{equation}\label{mu-omega13}
\mu_{ab}\simeq 2{\cal Q}_{a|1}^- {\cal Q}_{b|3}^-\omega^{13},
\end{equation}
where, importantly, $\omega^{13}$ is a periodic function, whereas ${\cal Q}_{a|i}^-$
is analytic in the upper half plane.
We denote the LO and NLO orders \(\omega^{13}=-\omega_{24}\simeq w^{-2}\,\omega^{13}_{(0)}+w^{-1} \,\omega^{13}_{(1)}\)
and \(\mu_{ab}\simeq w^{-2}\,\mu_{ab}^{(0)}+w^{-1} \,\mu_{ab}^{(1)}\) and represent \eqref{mu-omega13} in the form
\begin{equation}\label{mu/omega13}
\frac{\mu_{ab}}{\omega^{13}}\simeq\frac{\mu_{ab}^{(0)}}{\omega^{13}_{(0)}}\left(1+w\left(\frac{\mu_{ab}^{(1)}}{\mu_{ab}^{(0)}} -\frac{\omega^{13}_{(1)}}{\omega^{13}_{(0)}} \right)\right)\simeq
2{\cal Q}_{a|1}^- {\cal Q}_{b|3}^-.
\end{equation}
First, let us look at the leading term in the r.h.s.: $\frac{\mu_{ab}^{(0)}}{\omega^{13}_{(0)}}$
 should be analytic in the upper half plane and it has a power-like asymptotics, as  the r.h.s. does.
This means that $\omega_{(0)}^{13}=B \sinh^2(\pi u)$ and thus $\frac{\mu_{ab}^{(0)}}{\omega^{13}_{(0)}}\sim
\frac{\mu_{ab}^{(0)}}{\sinh^2(\pi u)}
=P^-_{ab}$ is a polynomial.
So \(\frac{\omega^{13}_{(1)}}{\omega^{13}_{(0)}}\)  could contain in the upper half plane a sum of poles  of the second order and of the first order at all \(u=in,\,\,n\in \mathbb{Z}\) with equal residues (because $\omega^{13}$ is periodic). To preserve the
analyticity in the upper half plane
we should cancel these poles by  the poles in \(\frac{\mu_{ab}^{(1)}}{\mu_{ab}^{(0)}}\).
Also we note that the ratio \(\frac{\mu_{ab}^{(1)}(u+i/2)}{\mu_{ab}^{(0)}(u+i/2)}\) is an even function, which also fixes the pole structure in the lower half plane.
In addition \(\frac{\mu_{ab}^{(1)}}{\mu_{ab}^{(0)}}\) could have a finite number of
poles at zeros of the polynomial $P_{ab}^-$.
In other words, the most general function with these properties can be written as
\beq
\frac{\mu_{ab}^{(1)}(u+i/2)}{\mu_{ab}^{(0)}(u+i/2)}=\frac{1}{\cosh^2(\pi u)}r_{ab}
+p_{ab}\Psi(u)+\frac{\CR_{ab}(u)}{P_{ab}(u)}
\eeq
where $r_{ab}$ and $p_{ab}$ are some constants and ${\cal R}_{ab}(u)$ are regular functions.
The first term represents an infinite series of the second order poles with equal residues
and the second term gives an infinite series of the first order poles with equal residues since
\begin{equation}\label{Psi-def}
\Psi(u)=\psi\left(\frac{1}{2}-iu\right)+\psi\left(\frac{1}{2}+iu\right)-2\psi(1)\;.
\end{equation}
The last term takes into account the possibility that there are extra poles canceled by the ratio $\frac{\mu_{ab}^{(0)}}{\omega^{13}_{(0)}}$ outside the brackets.
We notice that ${\cal R}_{ab}$ can only be a polynomial of the same order as $P_{ab}$.
Thus again we have a small number of constant coefficients in our ansatz to fix.

We can fix  \(r_{ab}\), \(p_{ab}\) and \(\CR_{ab}(u)\) in the same way as we did for the leading order
i.e. by applying \eqref{fthorder} and the regularity conditions,
telling that the combinations $\mu_{ab}+\tilde\mu_{ab}$
and $\frac{\mu_{ab}-\tilde\mu_{ab}}{\sqrt{u^2-4g^2}}$ are regular at $u\sim 0$.
This procedure leads to the following result
\begin{eqnarray*}
&&r_{ab}=2\pi^2\Lambda\,,\qquad p_{ab}=-\frac{1}{2}\,,  \\
&& \CR_{12}=\frac{1}{\pi^2 \Lambda^2}\frac{8i\left(2 \pi ^2 \left(\Delta ^2-1\right) \Lambda +3\right)}{3\left(\Delta ^2-1\right)^2}, \\
&& \CR_{13}=\frac{1}{\pi^2 \Lambda^2}\frac{3-2 \pi ^2 \left(\Delta ^2-1\right)\Lambda}{6}u, \\
&& \CR_{14}=-\frac{1}{\pi^2 \Lambda^2}\left(\frac{2\pi^2(\Delta^2-1)\Lambda+3}{48}(4u^2+1)-\frac{(\Delta^2-1)\Lambda}{3}\right), \\
&& \CR_{24}=-\frac{1}{\pi^2 \Lambda^2}\frac{ \left(2 \pi ^2 \left(\Delta ^2-1\right) \Lambda +9\right) }{72}u\left(4 u^2+1\right), \\
&& \CR_{34}=\frac{1}{\pi^2 \Lambda^2}\frac{i(\Delta^2-1)^2}{18432}(4u^2+1)((2\pi^2(\Delta^2-1)\Lambda+3)(4u^2-3)+72(\Delta^2-1)\Lambda).
\end{eqnarray*}
and also fixes the reminding coefficient
in \eq{Pmain} to \eq{c312}.

\section{NLO solution for $Q$}\la{AppQ}
By making the Mellin transformation of $Q(u)$
we converted the finite difference equation \eq{Baxter-w}
into a second order PDE which we managed to solve and transform the solution back
explicitly. The result we found reads
\small
\begin{multline}\label{QsolNLO}\nn
\frac{\sqrt{w}(u^2-2\Lambda w)}{-i u-\frac{w}{4}+i\sqrt{2\Lambda w}}{\Gamma \left(-iu+\frac{w}{4}+{i} \sqrt{2 \Lambda w}\right) \over \Gamma\left(-iu-\frac{w}{4}-i\sqrt{2 \Lambda w}\right)}
\,_3F_2\left(\frac{1-\Delta}{2},\frac{1+\Delta}{2},-i u-\frac{w- i\sqrt{32 \Lambda w} }{4};-\frac{w}{2},2i\sqrt{2 \Lambda w}+1;1\right).
\end{multline}
\normalsize
Note that this solution contains $\sqrt{w}$ terms. As the initial equation is analytic in $w$
changing the sign of $\sqrt{w}$ we get two linear independent solutions.
Suitable combinations of these two solutions
should give $\bQ_1$ and $\bQ_3$ with ${\cal O}(w)$ accuracy.
As this result is not required for the leading order calculation of this paper
these combinations will be published elsewhere.

\section{Finding $\omega_{ij}$}\label{app_omega}

Combining \eq{bQt} and \eq{Qt3} we can extract
\begin{align}
& \omega_{12}=-\cos\left(\frac{\pi\Delta}{2} \right)\;, \\
& \omega_{14}=-i\frac{\cosh\left(\pi\left(u+\frac{i\Delta}{2}\right) \right)}{\sinh(\pi u)}\;, \\
& \omega_{34}=-\frac{\cos\left(\frac{\pi\Delta}{2} \right)}{\sinh^2(\pi u)}-2i\tan\left(\frac{\pi\Delta}{2} \right)\frac{\cosh\left(\pi\left(u+\frac{i\Delta}{2}\right) \right)}{\sinh(\pi u)}\;.
\end{align}
Next, using the unit Pfaffian constraint we get
\begin{equation}
\omega_{12}\omega_{34}-\omega_{13}\omega_{24}+\omega_{14}^2=1\;,
\end{equation}
from where we obtain
\begin{equation}\la{constr}
\omega_{13}\omega_{24}=-2\;.
\end{equation}
As we discussed in Appendix~\ref{NLOP}
\begin{align}
& \omega_{24}=\frac{B\sinh^2(\pi u)}{w^2}\;,
\end{align}
i.e. from \eq{constr}
\beq\la{o13}
\omega_{13}=-\frac{2w^2}{B\sinh^2(\pi u)}\;.
\eeq
To complete the calculation we have to find the constant $B$.
We note that $B$ can be extracted from the singularity of $\omega_{13}$ which
can be computed independently from
\begin{equation}\label{method}
\omega_{13}=\frac{\omega_{13}+\tilde{\omega}_{13}}{2}+\frac{\omega_{13}-\tilde{\omega}_{13}}{2\sqrt{u^2-4\Lambda w}}\sqrt{u^2-4\Lambda w},
\end{equation}
where the combinations \(\frac{\omega_{13}+\tilde{\omega}_{13}}{2} \) and \(\frac{\omega_{13}-\tilde{\omega}_{13}}{2\sqrt{u^2-4\Lambda w}} \)
are regular around $u=0$.
We shell use that
\begin{equation}
\frac{\omega_{13}-\tilde{\omega}_{13}}{2\sqrt{u^2-4\Lambda w}}=
\frac{
\tilde{\bQ}_1 \bQ_3-\bQ_1 \tilde{\bQ}_3}{2\sqrt{u^2-4\Lambda w}}=\left(-\frac{16\cos\frac{\pi\Delta}{2}}{\pi^2(\Delta^2-1)^2}u+\mathcal{O}(u^3) \right)+\mathcal{O}(w).
\end{equation}
Therefore for the leading singularity of \(\omega_{13}\) we have
\begin{equation}
\omega_{13}=w^2\left(\frac{32\cos\frac{\pi\Delta}{2}}{\pi^2(\Delta^2-1)^2}\frac{\Lambda^2}{u^2}+\mathcal{O}\left(\frac{1}{u} \right)\right)
+
\mathcal{O}(w^3)
\end{equation}
comparing with \eq{o13} we get
\begin{equation}
B=-\frac{(\Delta^2-1)^2}{16\Lambda^2\cos\frac{\pi\Delta}{2}}\;.
\end{equation}

\bibliography{bibliography}
\bibliographystyle{JHEP}

\end{document}